\documentclass[10pt, conference]{IEEEtran}
\IEEEoverridecommandlockouts
\usepackage{cite}
\usepackage{amsmath,amssymb,amsfonts}
\usepackage{algorithmic}
\usepackage{graphicx}
\usepackage{textcomp}
\def\BibTeX{{\rm B\kern-.05em{\sc i\kern-.025em b}\kern-.08em
    T\kern-.1667em\lower.7ex\hbox{E}\kern-.125emX}}

\usepackage{marvosym}
\usepackage{makecell}
\usepackage{booktabs}  %% 三线表
\usepackage{multirow}  %%提供跨行命令\multirow{}{}{}
\usepackage{url}
\usepackage[backref]{hyperref}
\usepackage[ruled]{algorithm2e} %伪代码
\usepackage{wrapfig}   %%文字包裹图片
\usepackage{textcomp} % 引入textcomp包
\usepackage{pifont}
\usepackage{subfig}  % 加载 subfig 包
\usepackage{caption} % 加载 caption 包以便自定义图注样式
\usepackage[table]{xcolor}
\usepackage{enumitem}
\usepackage{tcolorbox}

\usepackage{tikz}
\usepackage{etoolbox}
\usepackage{enumitem}
\newcommand*{\circled}[1]{\lower.7ex\hbox{\tikz\draw (0pt, 0pt)%
    circle (.5em) node {\makebox[1em][c]{\small #1}};}}
\robustify{\circled}

\newtcolorbox[auto counter, number within=section]{mybox}[2][]{colback=gray!10!white, colframe=gray!70!black, boxrule=0.8mm, width=\linewidth, arc=1mm, rightrule=0mm, toprule=0mm, bottomrule=0mm, leftrule=1mm, #1}

\newcommand{\sysname}{{C-LLM}}
\newcommand{\funcname}{{SenteTruth}}
    
\begin{document}

% \title{Connecting Large Language Models with Blockchain: Making Smart Contracts Smarter}

\title{Connecting Large Language Models with Blockchain: Advancing the Evolution of Smart Contracts from Automation to Intelligence}

\author{Youquan Xian\IEEEauthorrefmark{1}\IEEEauthorrefmark{4}, Xueying Zeng\IEEEauthorrefmark{1}\IEEEauthorrefmark{4}, Duancheng Xuan\IEEEauthorrefmark{1}\IEEEauthorrefmark{4}, Danping Yang\IEEEauthorrefmark{1}\IEEEauthorrefmark{4}, Chunpei Li\IEEEauthorrefmark{1}\IEEEauthorrefmark{4}, Peng Fan\IEEEauthorrefmark{1}\IEEEauthorrefmark{4} and Peng Liu\textsuperscript{\Letter}\IEEEauthorrefmark{1}\IEEEauthorrefmark{4} \\
\IEEEauthorblockA{\IEEEauthorrefmark{1}\textit{Key Lab of Education Blockchain and Intelligent Technology, Ministry of Education, China} \\
\textit{\IEEEauthorrefmark{4}School of Computer Science and Engineering, Guangxi Normal University, China} \\
\textit{Corresponding author: Peng Liu (\tt\small liupeng@gxnu.edu.cn)}
}}

\maketitle

\begin{abstract}
Blockchain smart contracts have catalyzed the development of decentralized applications across various domains, including decentralized finance. However, due to constraints in computational resources and the prevalence of data silos, current smart contracts face significant challenges in fully leveraging the powerful capabilities of Large Language Models (LLMs) for tasks such as intelligent analysis and reasoning. To address this gap, this paper proposes and implements a universal framework for integrating LLMs with blockchain data, {\sysname}, effectively overcoming the interoperability barriers between blockchain and LLMs. By combining semantic relatedness with truth discovery methods, we introduce an innovative data aggregation approach, {\funcname}, which significantly enhances the accuracy and trustworthiness of data generated by LLMs. To validate the framework's effectiveness, we construct a dataset consisting of three types of questions, capturing Q\&A interactions between 10 oracle nodes and 5 LLM models. Experimental results demonstrate that, even with 40\% malicious nodes, the proposed solution improves data accuracy by an average of 17.74\% compared to the optimal baseline. This research not only provides an innovative solution for the intelligent enhancement of smart contracts but also highlights the potential for deep integration between LLMs and blockchain technology, paving the way for more intelligent and complex applications of smart contracts in the future.
\end{abstract}

% text-type
% semantic relatedness
% 公式里增加更新后的归一化过程,防止归0
% 总结
% table的攻击 名词


\section{INTRODUCTION}
\label{intro}
Blockchain smart contracts have been widely applied in fields such as decentralized finance \cite{schueffel2021defi}, supply chain management \cite{dutta2020blockchain}, and energy management \cite{wang2020integrating}, enabling the automatic execution of predefined rules to ensure transparency and security in transactions. However, current smart contracts remain in the automation stage, lacking intelligent decision-making capabilities, particularly in terms of semantic understanding and reasoning, which makes them ill-equipped to handle complex and dynamically changing scenarios.

\begin{figure}[hbtp]
  \centering
  \includegraphics[width=\linewidth]{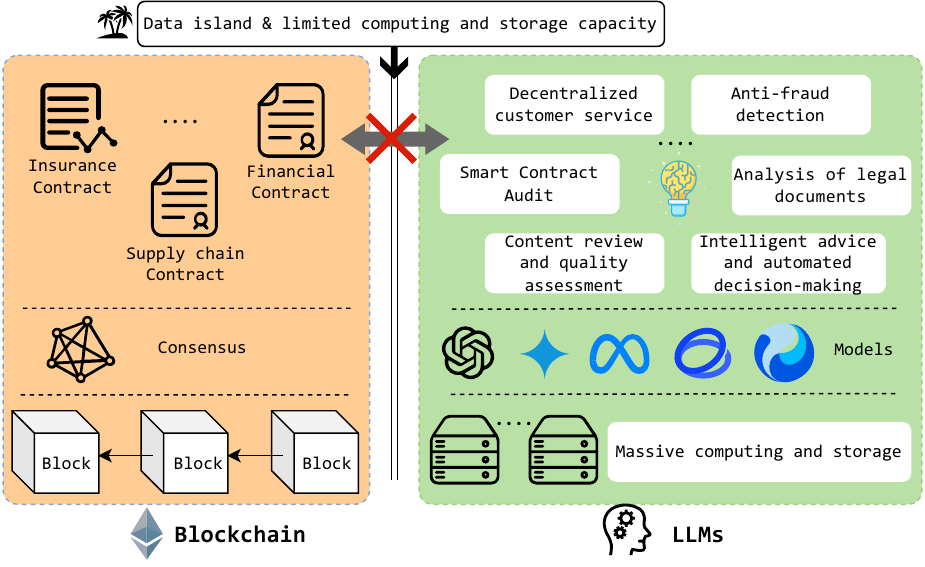}
  \caption{Obstacles to the intelligentization of smart contracts.}
  \label{fig:intro}
\end{figure}

Although blockchain technology offers unique advantages in terms of decentralization and transparency, its limited computational and storage resources hinder its ability to support high-performance Artificial Intelligence (AI) tasks, such as Large Language Models (LLMs). Moreover, the issue of data silos prevents seamless integration between blockchain and existing LLM services \cite{ren2023interoperability, belchior2021survey}, further restricting the intelligent evolution of smart contracts. Despite the powerful capabilities of LLMs in areas such as intelligent analysis, reasoning, intelligent assistants \cite{zhuang2023toolqa}, code understanding \cite{nam2024using}, and the detection and repair of smart contract vulnerabilities \cite{ma2024combining, luo2024fellmvp}, the data barriers between blockchain and LLMs remain a critical bottleneck for the intelligent enhancement of smart contracts.
Therefore, as shown in Fig. \ref{fig:intro}, breaking through the limitations of data silos and resolving the data interaction challenges between blockchain and LLMs are essential steps toward unlocking the full potential of smart contract intelligence. Such advancements will not only enhance the decision-making capabilities of smart contracts but also drive the broader adoption of blockchain in more complex, dynamic application scenarios.

\begin{figure}[hbtp]
  \centering
  \includegraphics[width=\linewidth]{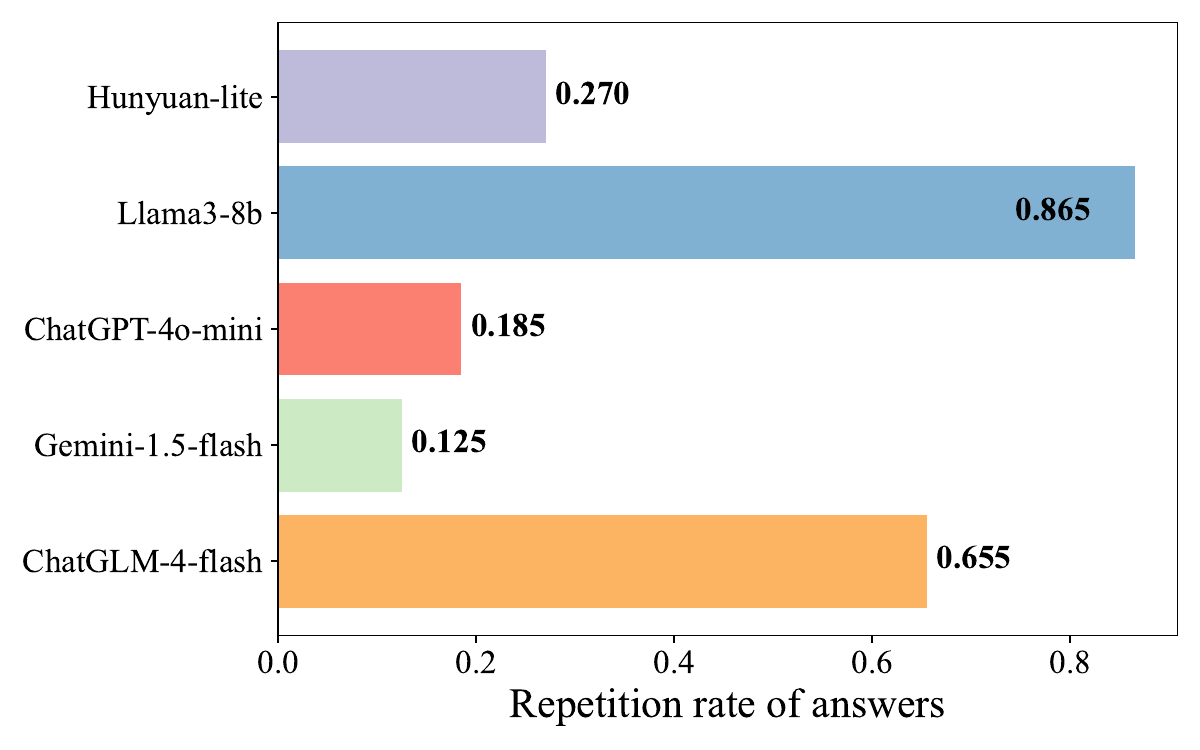}
  \caption{The repetition rate of answer indicates the highest proportion of identical answers generated by the LLM for the same question when LLM API request parameters were set as ``\texttt{temperature=0}, \texttt{top\_p=0}, \texttt{seed=42}''.}
  \label{fig:why1}
\end{figure}

To address this challenge, Xu et al. \cite{xu2023smartllm} proposed an initial concept of connecting blockchain and LLMs via oracles. As a trusted data middleware, an oracle ensures the integrity and trustworthiness of the data transmitted to the blockchain by leveraging technologies such as voting games \cite{peterson2015augur, berryhill2019astraea, gennaro2022deepthought}, trusted execution environments \cite{zhang2016town, liu2022extending, woo2020distributed}, and threshold signatures \cite{xian_iot, xian_iiot, dos}. These technologies help prevent data tampering during transmission, thus ensuring the authenticity and reliability of the data.
However, as illustrated in Fig. \ref{fig:why1}, we found that during the interaction between the oracle and the LLM, even when parameters such as ``\emph{temperature}'' and ``\emph{seed}'' are set in the LLM API request, complete consistency of the generated results cannot be guaranteed \cite{openai_api}. Furthermore, while adjusting request parameters to enhance the consistency of LLM-generated data often sacrifices the diversity and creativity of the generated content, the default settings of LLMs do not recommend such an approach, as they prioritize ensuring diversity and innovation in the output content.

However, this phenomenon results in inconsistencies in the data obtained by nodes, which in turn impairs the ability of oracle nodes to reach a consensus on data acquisition. Traditionally, oracle systems aggregate heterogeneous data using methods such as majority voting \cite{dong2023daon, chainlink}, median aggregation \cite{SchellingCoin, Oracul, Compound}, and truth discovery \cite{xiao2023decentralized, xian2024safeguarding, gigli2023decentralized}. However, these conventional methods are not well-suited for aggregating textual data returned by LLM APIs. Therefore, breaking down the data barriers between blockchain and LLMs, and addressing the interoperability challenges between LLMs and blockchain, becomes the central research objective of this paper.

We implement a general framework for LLM and blockchain data collaboration, {\sysname}, which effectively addresses the interoperability challenges between LLMs and blockchain. Even in the presence of inherent data inconsistencies and potential tampering attacks on the nodes, {\sysname} ensures the reliability of the data. Specifically, to tackle the issue of inconsistency in LLM-generated data, we design a novel data aggregation method, SenteTruth, which combines semantic relatedness with truth discovery techniques. This method significantly enhances the accuracy and trustworthiness of data obtained from large models. Furthermore, based on this framework, we construct a dataset consisting of three types of questions, including question-answer records from interactions between 10 nodes and five LLM models.

The main contributions are as follows: 
\begin{itemize} 
\item We implement a general framework for blockchain and LLM data collaboration, {\sysname}, which addresses the interoperability issues between LLMs and blockchain. The code and data are available at: \url{https://anonymous.4open.science/r/CLLM-5504}. 
\item We design a novel data aggregation method, SenteTruth, which combines semantic relatedness with truth discovery techniques to enhance the accuracy and trustworthiness of data obtained from large models. 
\item We analyze the effectiveness of the proposed scheme through experiments. In the presence of 40\% malicious nodes, the proposed method achieves an average improvement of 17.74\% in data accuracy compared to the optimal baseline. 
\end{itemize}

The remainder of this paper is structured as follows: Section \ref{bg} introduces the relevant background. Section \ref{main} presents the system flow and details of the proposed solution. Section \ref{experiment} provides the experimental results and analysis. Finally, Section \ref{conclusion} concludes the paper and discusses future work.

\section{RELATED WORK}
\label{bg}
To address the issue of data trustworthiness in oracle systems, Augur \cite{peterson2015augur} and Astraea \cite{berryhill2019astraea} employ a betting mechanism where multiple oracle nodes stake on the authenticity of the data, using a value-based game theory approach to incentivize honest behavior. Deepthought \cite{gennaro2022deepthought}, on the other hand, combines voting and reputation mechanisms to reduce the risk of corruption caused by adversarial nodes or inactive voters, thereby improving data reliability.
In another approach, Zhang et al. \cite{zhang2016town} and Liu et al. \cite{liu2022extending} integrated Trusted Execution Environment (TEE) technology with oracles to ensure the integrity of the data. Additionally, threshold signatures \cite{dos,lin2022novel} and improved TLS protocols, such as TLS-N \cite{zhang2020deco,luo2024proxying}, utilize cryptographic algorithms to guarantee data reliability, offering higher security compared to other methods. However, the implicit assumption of all these solutions is that the data obtained by different nodes is consistent. If the data retrieved by nodes is inconsistent, achieving data consensus becomes a challenging task.

To improve data consistency before reaching consensus, Liu et al. \cite{xian_iiot} utilized a sliding window mechanism to enhance the centralization of IoT real-time data before consensus is achieved. Xian et al. \cite{xian2024instant} proposed the representative enhancement aggregation strategy REP-AG and access timing optimization strategy TIM-OPT to improve data consistency for oracle nodes when acquiring real-time data.
In addition, selecting the median from the set of data obtained by nodes as the final answer is a common method for on-chain data consensus \cite{SchellingCoin,Oracul,Compound}. For off-chain consensus, DAON \cite{dong2023daon} uses methods such as majority voting or averaging to eliminate data inconsistencies before consensus is reached.
Xiao \cite{xiao2023decentralized} and Xian \cite{xian2024safeguarding} applied Truth Discovery (TD) \cite{li2016survey} to weigh and aggregate numerical data from different nodes based on their trustworthiness. They also reverse-update trustworthiness according to data deviations, thus making the aggregated data approach the ``truth''. Furthermore, sharing trustworthiness allows honest nodes to reach consensus on heterogeneous data. Similarly, Gigli et al. \cite{gigli2023decentralized} employed the Truth Inference algorithm to achieve consensus on numerical data from IoT sensors before formal consensus.

Despite the extensive research conducted, the aforementioned methods remain primarily applicable to numerical data types. When the data retrieved by the oracle is textual, such as that generated by the LLM API, these methods are unlikely to be effective. The inherent inconsistency in LLM-generated data further complicates the detection of malicious data manipulation by adversarial nodes within oracle systems. Therefore, the main research goal of this paper is to design an aggregation method suitable for text-type data for the oracle system, reduce the impact of malicious nodes tampering with data, and improve the reliability of the final data.

\begin{figure*}[t!]
  \centering
  \includegraphics[width=\linewidth]{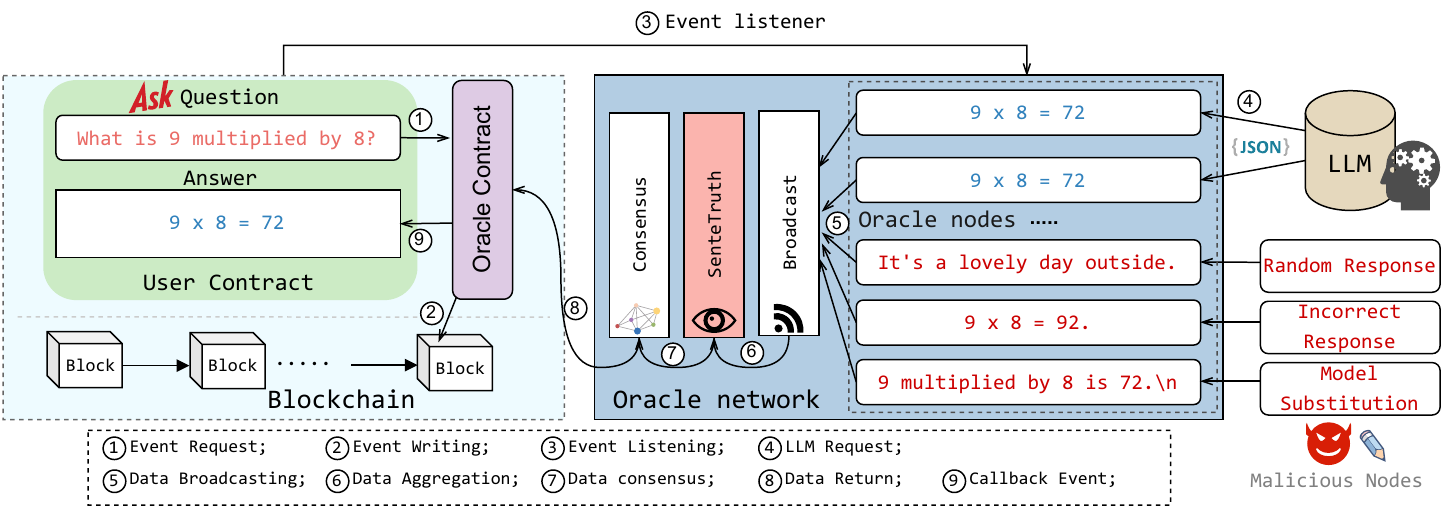}
  \caption{The system flow of \sysname.}
  \label{fig:overview}
\end{figure*}

\section{DESIGN OF \sysname}
\label{main}

As introduced in Section \ref{intro}, {\sysname} is a general-purpose data integration framework designed to provide trusted LLM data for blockchain, enabling secure interaction between the two. As shown in Fig. \ref{fig:overview}, it consists of the following four components:

\begin{itemize} 
\item \textbf{User Contract}: A smart contract, created and deployed by users within a blockchain network, is designed to automate and ensure the transparency of transactions and protocol execution. For example, in decentralized insurance applications, the smart contract automatically assesses claims based on predefined rules and executes payments without relying on traditional insurance companies as intermediaries. By leveraging smart contracts, insurance processes become transparent and immutable, significantly improving efficiency and reducing operational costs. In such scenarios, LLMs can be employed to automate document review, customer interaction, and intelligent reasoning in claims processing. These models can provide personalized insurance recommendations and claim suggestions, further enhancing the contract's intelligence.

\item \textbf{Oracle Contract}: A smart contract deployed on the blockchain is specifically designed to provide the necessary external data for the blockchain. Since blockchains cannot directly access off-chain information, oracle contracts act as intermediaries by acquiring external data (such as financial market prices, weather conditions, IoT sensor data, etc.), helping smart contracts make decisions. These contracts ensure that on-chain smart contracts can interact with real-world data,  thus making the application of decentralized applications more extensive.

\item \textbf{Oracle Network}: A decentralized network comprising multiple independent oracle nodes is designed to ensure the accuracy, reliability, and tamper resistance of data. The oracle network acquires and verifies external data from various nodes, transmitting the consensus data to the smart contracts on the blockchain. Unlike a single oracle, an oracle network mitigates the risks of data provider centralization, thereby preventing malicious attacks or the injection of erroneous information. Decentralized oracle networks, such as Chainlink \cite{chainlink} and DOS Network \cite{dos}, uphold data credibility through incentive mechanisms and consensus protocols. These networks are widely deployed in decentralized finance (DeFi), insurance, supply chain management, and other sectors.

\item \textbf{Large Language Models (LLMs)}: Natural language processing models with vast numbers of parameters are capable of understanding and generating complex linguistic content. Common LLMs include OpenAI’s ChatGPT\footnote{\url{https://chatgpt.com/}} , Google’s Gemini\footnote{\url{https://gemini.google.com/}} , and Meta’s Llama\footnote{\url{https://www.llama.com/}}. These models exhibit strong capabilities in tasks such as automated customer service, legal document review, medical diagnosis, and financial analysis. However, due to their large parameter sizes and complex training processes, LLMs require substantial computational resources and distributed computing platforms to operate efficiently.
\end{itemize}

\subsection{System Flow}
As shown in Fig. \ref{fig:overview}, the entire system process includes the following steps:

\begin{enumerate}[label=\circled{\arabic*}] 
\item Event Request: When the user contract needs to rely on the intelligent capabilities of the LLM to process a task, it invokes the oracle contract interface to initiate a task request $\mathbb{R}$. $\mathbb{R}$ contains information such as the task identifier $\mathcal{I}$, the question $\mathcal{Q}$, and the requested model $\mathcal{M}$. Specifically, $\mathcal{I}$ is the unique identifier of the task, used to track the task’s status and progress; $\mathcal{Q}$ is the question the user wants the LLM to answer or process, involving natural language processing, reasoning, and other tasks; $\mathcal{M}$ specifies the LLM model to be used, such as ChatGPT-4 or Llama-8b, to ensure the task is effectively completed.

\item Event Writing: After receiving the event request $\mathbb{R}$, the oracle contract writes the corresponding event record $\mathbb{E}$ into the blockchain and broadcasts it to the oracle network. The oracle contract typically also includes functions like node registration and payment, which are not discussed here.

\item Event Listening: Nodes $\mathcal{O}_i$ in the oracle network continuously listen for blockchain data request events $\mathbb{E}$ to respond to the user contract’s needs promptly.

\item LLM Request: When a node $\mathcal{O}_i$ detects event $\mathbb{E}$, it triggers an API request to the corresponding model $\mathcal{M}$ as specified by the event and processes the returned data result $\mathcal{D}_i$. It is important to note that some malicious nodes may obtain incorrect results $\mathcal{D}_i^{'}$ at this stage, as detailed in Section \ref{malicious}.

\item Data Broadcasting: Nodes in the oracle network exchange the data $\mathcal{D}_i$ they have obtained to ensure data consistency. However, to prevent the “Freeloading” issue \cite{chainlink}, nodes need to go through a two-round data exchange process (commit-reveal) to ensure the authenticity and integrity of the data. First, node $\mathcal{O}_i$ broadcasts the data hash $(\mathcal{I}, \text{Hash}(\mathcal{D}_i))$, which only transmits the hash value of the data, ensuring that other nodes cannot alter the data. Then, after collecting sufficient hashes, node $\mathcal{O}_i$ will broadcast its actual data $(\mathcal{I}, \mathcal{D}_i)$ to verify the data’s integrity and consistency. This process effectively prevents “Freeloading” behavior during data exchange.

\item Data Aggregation: After data broadcasting, each node will possess the data set $\mathcal{D}$ from all nodes. By running the SenteTruth algorithm, nodes can extract the truth $\bar{\mathcal{D}}$ from the disordered data, avoiding the influence of erroneous data from malicious nodes, as discussed in Section \ref{malicious}. The details of the algorithm are described in Section \ref{detail}.

\item Data Consensus: The nodes reach consensus on $\bar{\mathcal{D}}$. In terms of the consensus protocol, in addition to common algorithms like PBFT, threshold signature algorithms are also an effective method for achieving consensus in the oracle network. Threshold signatures allow multiple nodes to jointly sign data, and only when a sufficient number of nodes agree can a valid signature be generated.

\item Data Return: After reaching consensus, the final result $\tilde{\mathcal{D}}$ is returned to the oracle contract by node $\mathcal{O}_i$ through a data upload interface.

\item Callback Event: After the oracle contract verifies the result, it calls the callback function to return the final result $\tilde{\mathcal{D}}$ to the user contract and continues executing the subsequent program logic.
\end{enumerate}

\subsection{Adversary Model} 
\label{malicious} 
Considering the structural and characteristic differences between the text data generated by LLMs and the numerical data handled by previous applications like DAON \cite{dong2023daon}, we design three potential text data manipulation attacks based on the gaps in existing oracle research and an analysis of the potential risks associated with LLM-generated text data.
\begin{enumerate} 
\item Random Response: Nodes, due to lack of access to LLM APIs or to save on data access costs, may return a random, meaningless statement regardless of the query made. 
\item Model Substitution: Similar to the random response scenario, nodes may engage in model substitution due to the varying pricing of LLM APIs. They might opt to query a cheaper model for the same task, thereby profiting from the price difference. 
\item Incorrect Response: Malicious nodes can exploit prompt engineering techniques to intentionally elicit incorrect answers, thereby undermining the reliability of the oracle system \cite{zhao2024towards}. 
\end{enumerate}

However, we must also assume that the number of malicious nodes in the system can not be more than half and not collusion, which is the basic assumption of system security.

\subsection{SenteTruth}
\label{detail}

To address the issue of inconsistent data obtained by different nodes through LLM APIs and the potential data manipulation by malicious nodes in the oracle network, which may undermine the system's reliability, we propose a data aggregation scheme, SenteTruth, based on semantic relatedness and truth discovery techniques. This scheme analyzes the semantic relatedness of the data provided by different nodes, cross-validates it with node credibility, and mitigates the influence of malicious nodes, ensuring the reliability of the data.

To mitigate the impact of malicious content, we first assess the similarity of the data obtained by different nodes. We use the Sentence-BERT (SBERT) model, which maps text to a fixed-length vector space using a pre-trained BERT model \cite{reimers2019sentence}. These embedding vectors effectively capture sentence-level semantic information, overcoming the limitations of traditional word vector models in assessing semantic relatedness between sentences. Let the input text be represented as $\mathcal{D} = \{\mathcal{D}_1, \mathcal{D}_2, \dots, \mathcal{D}_n\}$, where $\mathcal{D}_i$ denotes the text returned by node $\mathcal{O}_i$.

\begin{eqnarray}
v_i = \text{SBERT}(\mathcal{D}_i)
\end{eqnarray}

Let $v_i$ represent the embedding vector of the answer $\mathcal{D}_i$ generated by the SBERT model. In this manner, each answer $\mathcal{D}_i$ is mapped to a fixed-length vector space, capturing its semantic information.

Once the text is encoded as vectors, we use cosine similarity to measure the relatedness between the data obtained by different nodes. Through this process, we can identify and filter out noise data that deviates significantly from the majority of the data, thereby reducing the impact of malicious nodes on the system's reliability.

\begin{eqnarray}
\varphi(v_i) = \sum_{j=1, j \neq i}^{n} \frac{v_i \cdot v_j}{\|v_i\| \|v_j\|}
\end{eqnarray}

However, for some open questions, LLM responses may exhibit significant variability, which makes semantic relatedness-based measures insufficient to accurately reflect the truth of the data. Therefore, to improve the accuracy of these data assessments, we integrate truth discovery \cite{li2016survey} into the process.

\begin{figure}[h]
  \centering
  \includegraphics[width=\linewidth]{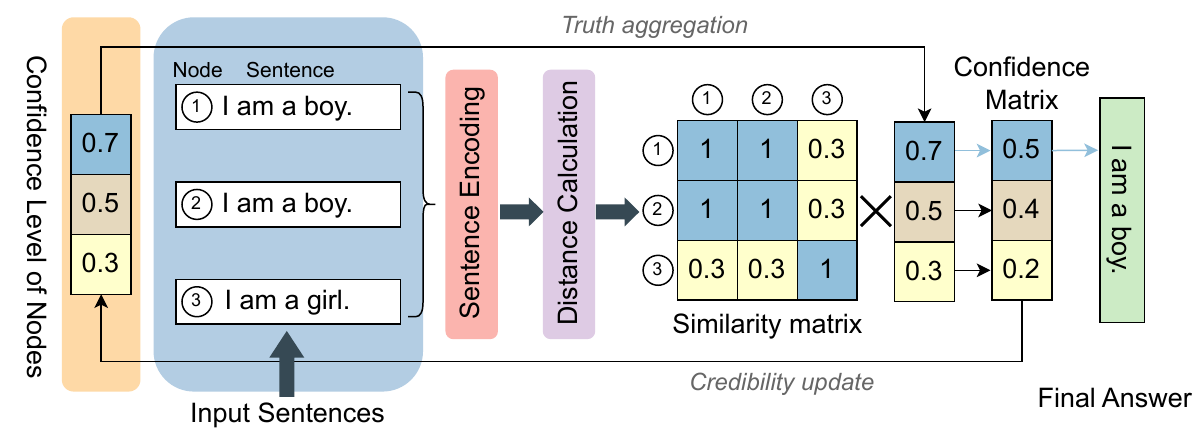}
  \caption{Details of \funcname.}
  \label{fig:deltailed}
\end{figure}

First, truth discovery can be divided into two main stages: truth aggregation and credibility updating. In the truth aggregation stage, the goal is to synthesize an aggregated result that approximates the true value as closely as possible, based on the data provided by multiple nodes. Here, the truth refers to the data that reflects the majority consensus of the nodes and is free from tampering, even if the response is not entirely accurate in content. To enhance the accuracy of the aggregation, we treat the data provided by each node as a potential hypothesis and weight these hypotheses based on both semantic relatedness and the credibility of the nodes. The credibility of node $\mathcal{O}_i$, denoted as $C_i$, reflects its historical performance. A higher credibility $C_i$ indicates that the data provided by $\mathcal{O}_i$ in past tasks was more consistent with the data provided by the majority of nodes.
When performing truth aggregation, we calculate the aggregated result by weighting both the credibility and the text similarity as follows:

\begin{eqnarray}
\bar{\mathcal{D}}  \gets  \text{argmax}_{i} C_i  \cdot \varphi(v_i)
\end{eqnarray}

In the credibility update phase, we can update the credibility \( C_i \) of a node based on the semantic relatedness between the data provided by the node and the aggregation result. Specifically, if the data provided by the node is similar to the aggregation result, i.e., similar to the data of other nodes, the node's credibility increases; otherwise, it decreases. Since \( \varphi(v_i) < 1 \), in order to avoid excessive fluctuations in the credibility calculation, we use the following formula for credibility update:

\begin{eqnarray}
C_i \gets \frac{\sum_{i=1}^n{C_i}}{\sum_{i=1}^n{C_i  \cdot \varphi(v_i)}} \cdot C_i  \cdot \varphi(v_i)
\end{eqnarray}

This improvement ensures that the truth discovery method can also effectively support text-based data returned by LLMs.

\section{PERFORMANCE EVALUATION}
\label{experiment}

\subsection{Experimental Setup}
To validate the effectiveness of the proposed scheme, we constructed a local Ethereum environment based on Ganache\footnote{\url{https://archive.trufflesuite.com/ganache/}}, implemented smart contracts using Solidity, and facilitated communication between the oracle nodes and the blockchain through Web3.py. The oracle system includes a user contract, an oracle contract, and an oracle network composed of 10 oracle nodes. Additionally, to ensure the verifiability and generalizability of the experimental results, we designed a local dataset containing questions and answers posed to five large language models (ChatGPT-4o-mini, Gemini-1.5-flash, Llama3-8b, ChatGLM-4-flash, Hunyuan-lite) by 10 different nodes. We used default parameters and recorded their responses. The dataset and code are available at \url{https://anonymous.4open.science/r/CLLM-5504}.

Considering that traditional aggregation methods are generally focused on numerical data and have limited effectiveness when dealing with text data, we established several baseline methods for text data aggregation for comparison. First, the classic majority voting method \cite{dong2023daon,chainlink} was used as the baseline aggregation strategy. Second, to evaluate the performance of text similarity-based aggregation, we introduced two commonly used text encoding methods: TF-IDF \cite{qaiser2018text} and SBERT \cite{reimers2019sentence}. These two methods provide more precise support for calculating text similarity, thereby improving the accuracy of data aggregation.

\begin{figure*}[t!]
\centering
\subfloat[BASE Chinese]{\includegraphics[width=0.3\linewidth]{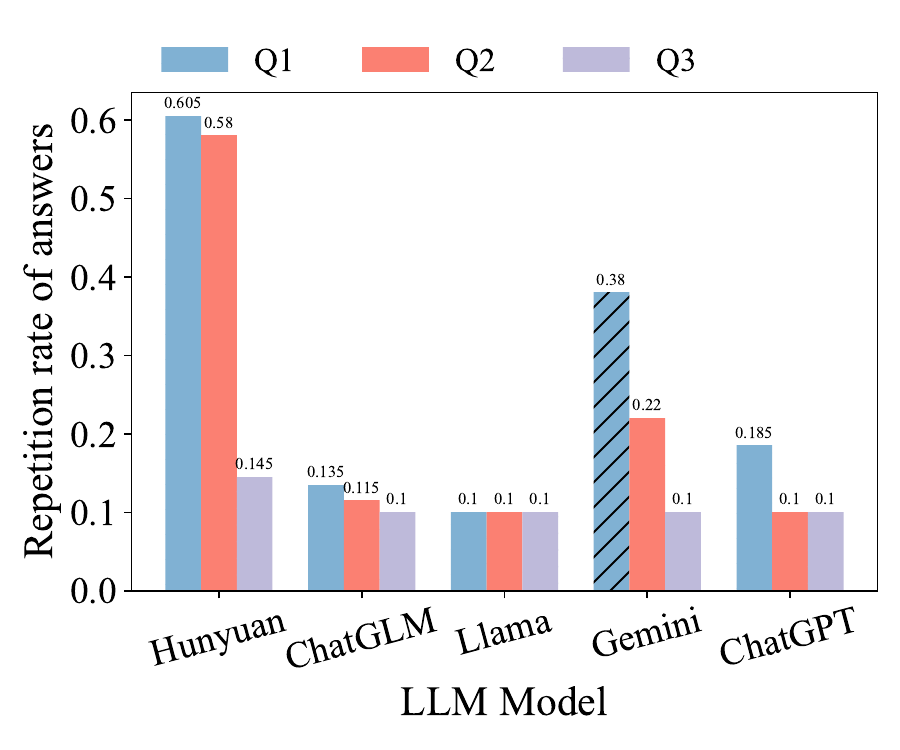}%
\label{fig:base_zh}}
\hfil
\subfloat[BASE English]{\includegraphics[width=0.3\linewidth]{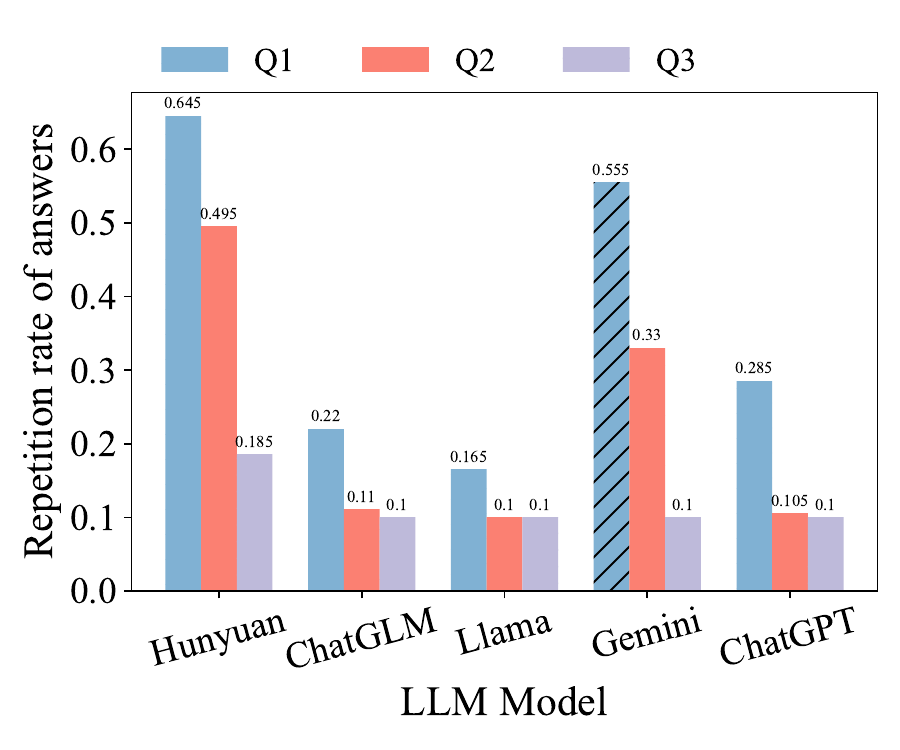}%
\label{fig:base_en}}
\hfil
\subfloat[MIX]{\includegraphics[width=0.3\linewidth]{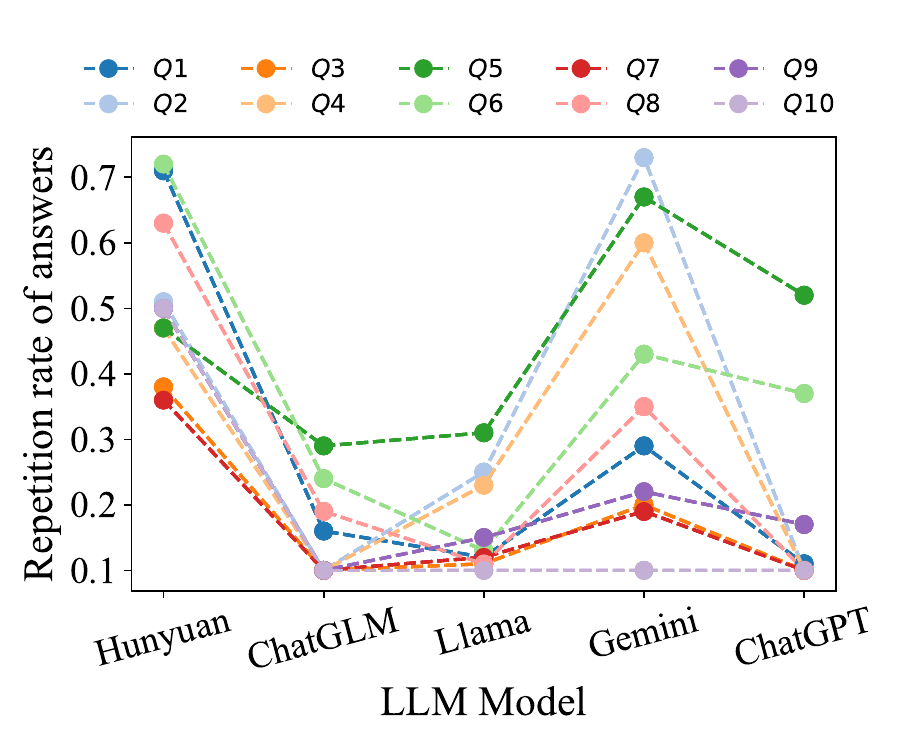}%
\label{fig:mix}}
\caption{The influence of different models and problem types on data consistency.}
\label{fig:data}
\end{figure*}

To comprehensively evaluate the proposed aggregation method, we constructed three different types of datasets, each testing the performance of the aggregation algorithms from various perspectives: basic questions, complex scenarios, and specialized domains.
\begin{enumerate} 
\item \textbf{Base Dataset (BASE)}: This dataset contains three types of questions, with 20 questions in each category, designed to evaluate the effectiveness of the aggregation methods across different question types: 
\begin{itemize} 
\item \textbf{Fact Consistency Questions (Q1)}: These questions assess the ability of the aggregation method to identify and exclude factually inconsistent data. For example: \emph{``Please list the eight planets in the solar system.''}          
\item \textbf{Logical Consistency Questions (Q2)}: These questions are used to test the aggregation method's performance in identifying and eliminating logically inconsistent data. For example: \emph{``If a number is even, is twice that number also even?''} 
\item \textbf{Open Questions (Q3)}: This category aims to test how the aggregation method handles open-ended questions, particularly how it identifies and eliminates outlier data. For example: \emph{``If you could redesign the education system, how would you improve it?''} \end{itemize} 
\item \textbf{Mixed Dataset (MIX)}: This dataset consists of 100 questions across 10 different fields such as history, mathematics, and others. It is designed to simulate the diversity of real-world questions and validate the robustness of the proposed method when handling complex problems. 
\item \textbf{Professional Dataset (PRO)}: Based on the C-Eveal dataset \cite{huang2023ceval}, this set includes 20 physics test questions each from middle school, high school, and university levels. By adjusting the prompt (e.g., asking for only the answer or the answer with an explanation), we verify the effectiveness of the aggregation method in different question-answering scenarios. 
\end{enumerate}

\begin{figure}[h]
\centering
\subfloat[temperature]{\includegraphics[width=0.45\linewidth]{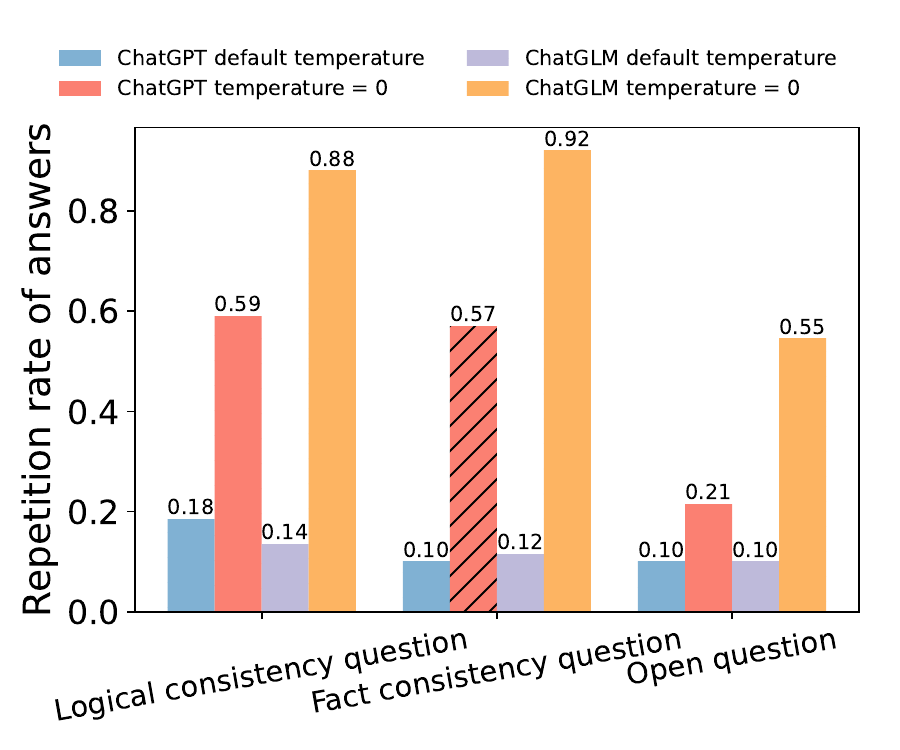}%
\label{fig:why_temperature}}
\hfil
\subfloat[prompt]{\includegraphics[width=0.45\linewidth]{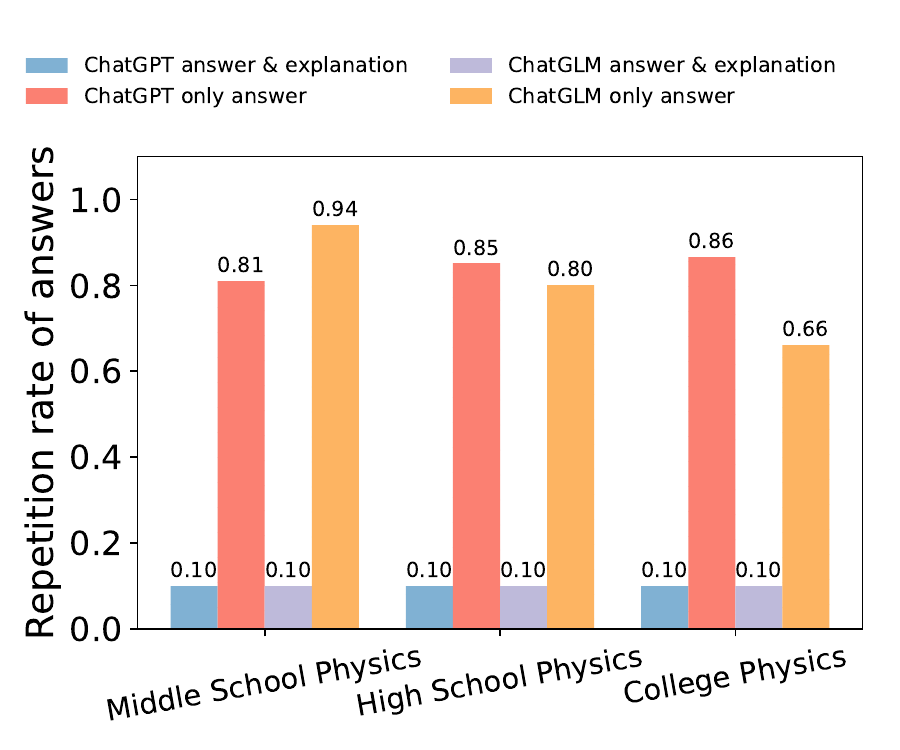}%
\label{fig:why_prompt}}
\caption{The influence of LLM API temperature parameters and Prompt on data consistency. The default temperature of ChatGPT is 1.0, and the default temperature of chatGLM is 0.7.}
\label{fig:why}
\end{figure}

\subsection{Research Questions (RQs)}

In oracle systems interacting with large language models (LLMs), data inconsistency often arises, which not only affects the accuracy of the data but also complicates the detection of malicious behaviors. To address this issue, we designed a series of experiments aimed at analyzing the key factors affecting LLM response consistency and evaluating the effectiveness and practicality of the proposed method in improving data correctness.

These experiments aim to answer the following research questions (RQs):

\begin{enumerate} 
\item \textbf{RQ1 - Problem Analysis}: What factors contribute to the data inconsistency in LLM responses? This question aims to identify the key factors influencing LLM response consistency, providing a basis for the design of subsequent solutions. 
\item \textbf{RQ2 - Performance Comparison}: How does the proposed method perform in improving data correctness, and what advantages does it have compared to existing data aggregation methods? This question will evaluate the effectiveness of the proposed method in ensuring data Accuracy, with data Accuracy defined as the proportion of unaltered data in the final aggregated result. 
\item \textbf{RQ3 - Effectiveness Analysis}: Why does the proposed method effectively improve data correctness? This question aims to explore the underlying reasons for the effectiveness of the proposed method and analyze its mechanisms for enhancing data correctness. 
\item \textbf{RQ4 - Usability Analysis}: Are there any resource overheads in the proposed method that could hinder its practical application? This question will assess the usability and scalability of the proposed method. 
\end{enumerate}

\subsection{RQ1 - Problem Analysis}

As shown in Fig. \ref{fig:data}, the impact of different question types on data consistency obtained by oracle nodes is illustrated. By comparing Fig. \ref{fig:base_zh} and \ref{fig:base_en}, we observe that questions with a definitive answer exhibit higher data consistency compared to open-ended questions. Furthermore, the consistency variations across different question types in Fig. \ref{fig:mix} further validate this observation. Additionally, we notice significant differences in data consistency between various LLMs under the default request parameters, which could provide valuable guidance for model selection.

Fig. \ref{fig:why} demonstrates the influence of adjusting temperature parameters and modifying prompts in LLM APIs on data consistency. The results in Fig. \ref{fig:why_temperature} align with the official documentation of LLM APIs, showing that lowering the sampling temperature significantly reduces the randomness in the output. Fig. \ref{fig:why_prompt} further confirms the impact of controlling the prompt (e.g., requiring only the correct answer versus asking for both the correct answer and an explanation) on data consistency. By restricting the output to the correct answer, data consistency is significantly improved. However, in the PRO dataset, consisting of physics questions (middle school, high school, and university-level), the difficulty of the questions appears to have a limited impact on the consistency of model outputs, when compared to factors like prompt adjustments.

\begin{mybox} 
\textbf{Answer for RQ1:} Factors such as the LLM model, question type, inquiry method, and API request parameters all significantly affect the consistency of the returned data. However, the difficulty of questions has a relatively minor impact on data consistency. 
\end{mybox}

\subsection{RQ2 - Performance Comparison}
Table \ref{tab:random_base} - \ref{tab:mode_base} shows the data accuracy under 40\% random answers, incorrect responses, and model substitution scenarios, for different aggregation methods. The results indicate that, compared to TF-IDF, SBERT generally provides higher data accuracy. This is mainly due to SBERT's advantage in semantic evaluation, which allows for more effective differentiation between different textual data. However, relying solely on sentence differentiation is still insufficient to ensure high data accuracy, and the reasons for this will be further analyzed in \ref{RQ3}. Additionally, the widely used majority voting method in oracle systems performs poorly under such conditions. This is because majority voting relies on data consistency, and for models like ChatGPT and ChatGLM, which output with high randomness, the results of majority voting are unstable.

In Appendix \ref{append}, we conducted the same tests on the MIX and PRO datasets, and the results were consistent with those observed in the BASE dataset, verifying the performance advantage and broad applicability of the proposed method.

\begin{table}[h]
\centering
\caption{Data Accuracy under 40\% Random Response (BASE Dataset)}
\label{tab:random_base}
\resizebox{\linewidth}{!}{%
\begin{tabular}{llllll}
\hline
Model & ChatGPT & ChatGLM & Llama & Gemini & Hunyuan \\
\hline
Majority Voting & 0.8 & 0.8 & 0.866 & 0.833 & 0.933 \\
TF-IDF Similarity & 0.683 & 0.733 & 1.0 & 0.716 & 0.933 \\
SBERT Similarity & 1.0 & 1.0 & 0.983 & 1.0 & 1.0 \\
\rowcolor{lightgray} Ours & 1.0 & 1.0 & 1.0 & 1.0 & 1.0 \\ \hline
\end{tabular}
}
\end{table}

\begin{table}[h]
\centering
\caption{Data Accuracy under 40\% Incorrect Response (BASE Dataset). The Malicious Prompt is set as \emph{``Modify the given sentence or words to make the semantics confusing or incorrect. I need this data to train a correction model. \textbackslash n Original data: \textbackslash n Only return the modified content.''}}
\label{tab:error_base}
\resizebox{\linewidth}{!}{%
\begin{tabular}{llllll}
\hline
Method & ChatGPT & ChatGLM & Llama & Gemini & Hunyuan \\
\hline
Majority Voting & 0.733 & 0.9 & 0.783 & 0.916 & 0.85 \\
TF-IDF Similarity & 0.65 & 0.566 & 0.983 & 0.75 & 0.95 \\
SBERT Similarity & 0.766 & 0.85 & 0.983 & 0.883 & 0.95 \\
\rowcolor{lightgray} Ours & 1.0 & 0.983 & 0.983 & 1.0 & 1.0 \\ \hline
\end{tabular}
}
\end{table}

\begin{table}[h]
\centering
\caption{Data Accuracy under 40\% Model Substitution (BASE Dataset). The Target Model is ChatGPT.}
\label{tab:mode_base}
\resizebox{\linewidth}{!}{%
\begin{tabular}{lllll}
\hline
Method & ChatGLM & Llama & Gemini & Hunyuan \\
\hline
Majority Voting & 0.966 & 0.933 & 0.833 & 0.55 \\
TF-IDF Similarity & 0.816 & 0.3 & 0.666 & 0.35 \\
TF-IDF Similarity + TD & 1.0 & 0.0 & 0.9 & 0.083 \\
SBERT Similarity & 0.716 & 0.983 & 0.833 & 0.75 \\
\rowcolor{lightgray} Ours & 1.0 & 1.0 & 1.0 & 0.883 \\ \hline
\end{tabular}
}
\end{table}

\begin{figure*}[t!]
\centering
\subfloat[BASE]{\includegraphics[width=0.3\linewidth]{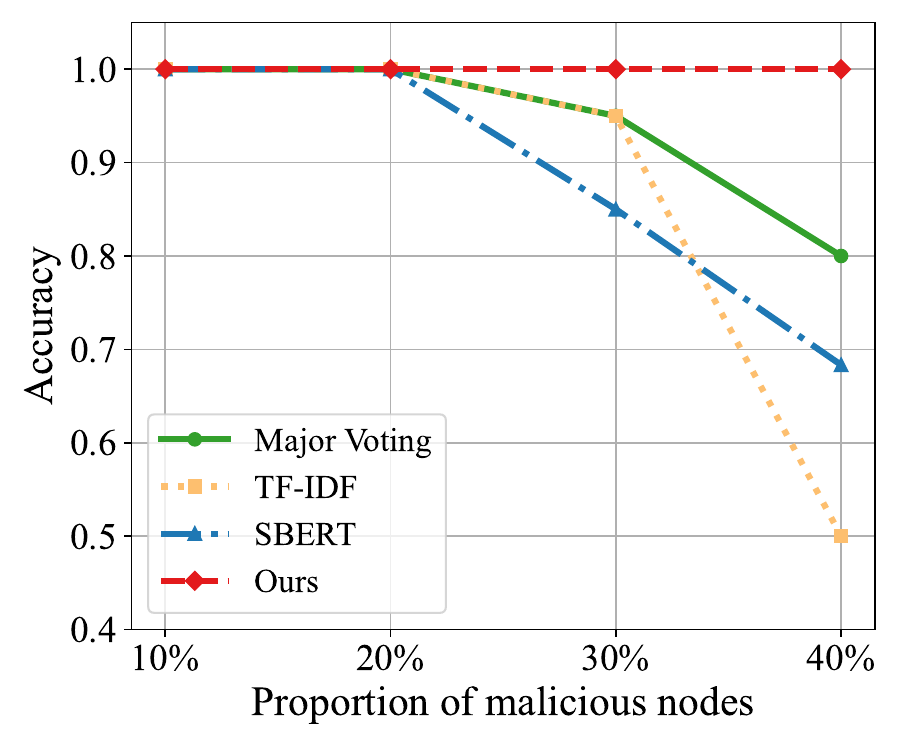}%
\label{fig:malicious_base}}
\hfil
\subfloat[MIX]{\includegraphics[width=0.3\linewidth]{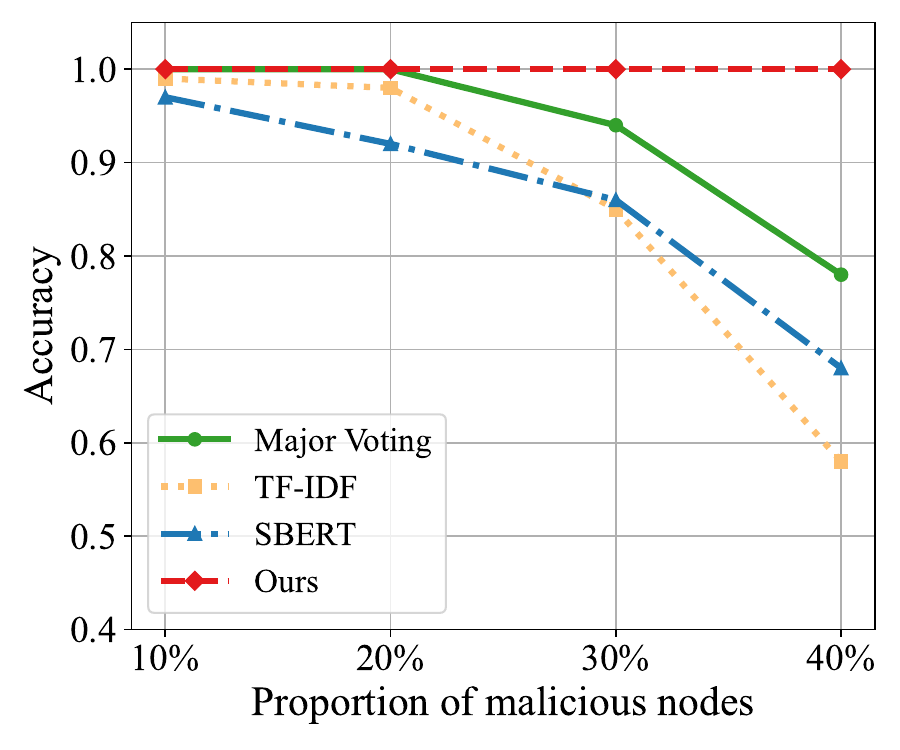}%
\label{fig:malicious_mix}}
\hfil
\subfloat[PRO]{\includegraphics[width=0.3\linewidth]{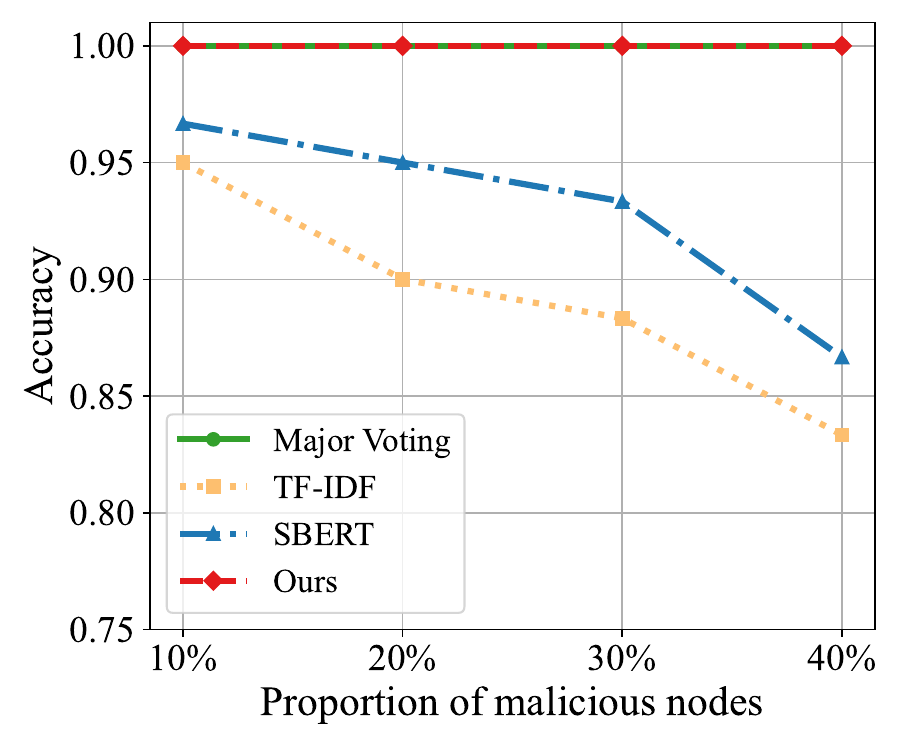}%
\label{fig:malicious_pro}}
\caption{The influence of Malicious Node Proportion on Data Accuracy.}
\label{fig:malicious}
\end{figure*}

\begin{figure*}[t!] 
\centering 
\includegraphics[width=\linewidth]{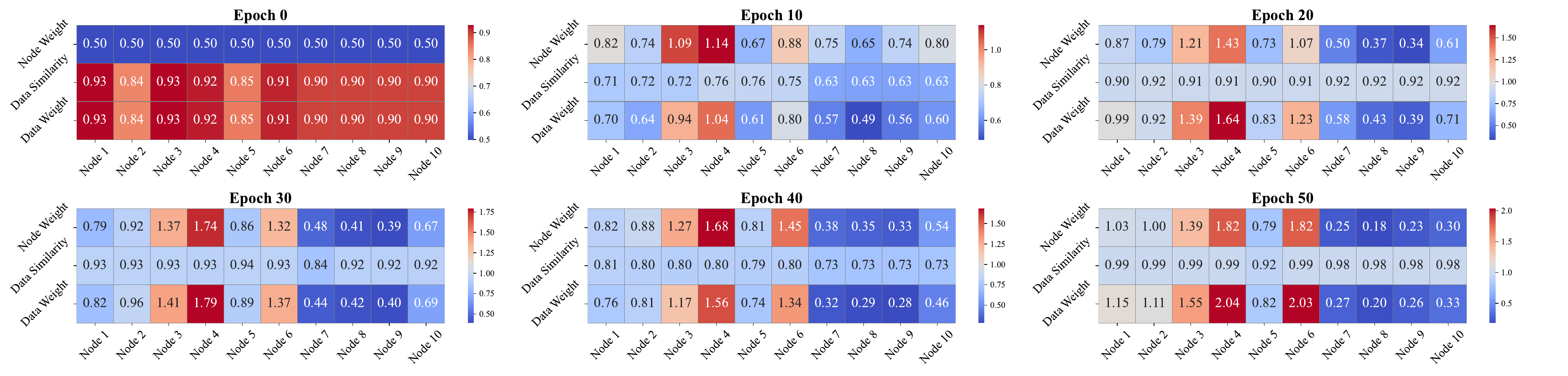} 
\caption{Key variable variations of the proposed method, including node weights, data similarity, and data weights. Nodes 7-10 are malicious nodes, which replace the original ChatGPT responses with Gemini responses.} 
\label{fig:hotmap} 
\end{figure*}

Fig. \ref{fig:malicious} analyzes the data accuracy under different proportions of malicious nodes. It can be observed that, with 40\% of malicious nodes, the proposed method improves data accuracy by an average of 17.74\% compared to the optimal baseline. Additionally, as the proportion of malicious nodes increases, methods based on SBERT, among others, show a significant drop in accuracy, while the proposed method still maintains a high level of accuracy. However, it is also noted that the majority voting method performs well in the PRO dataset, because in the PRO dataset, the answers only return the correct option, which ensures high data consistency.

\begin{mybox} \textbf{Answer for RQ2: } The proposed method demonstrates significant performance improvements over baseline methods. Even with different datasets and varying proportions of malicious nodes, the proposed method continues to maintain high data accuracy. \end{mybox}

\subsection{RQ3 - Effectiveness Analysis}
\label{RQ3}

Fig. \ref{fig:hotmap} illustrates the variation of key variables across different epochs in the BASE dataset. It can be observed that, at Epoch 0, the node weights are uniform; however, as the epoch progresses, the weights of the malicious nodes (Node 7-10) gradually decrease, indicating that SenteTruth performs effectively. However, as shown in Table \ref{tab:mode_base}, when the TD method is applied to TF-IDF, the final accuracy declines further due to TF-IDF's limited discriminative power. This suggests that using the truth discovery method to already discriminative text methods, such as SBERT, can further enhance data accuracy, and vice versa. Therefore, replacing SBERT with a stronger text encoding method is expected to continue improving the performance of SenteTruth.

Additionally, it is noteworthy that, for example, at Epoch 20, even if malicious nodes have a higher data similarity than honest nodes due to various reasons, their lower node weights from prior malicious behavior ensure that their tampered data is not selected in the final aggregation. This is both a limitation of SBERT and the reason why SenteTruth achieves higher accuracy—errors in distinguishing by SBERT are corrected due to the accumulated weight of correctly identified data from the majority.

\begin{mybox} 
\textbf{Answer for RQ3: } SenteTruth mitigates the impact of errors during data aggregation by accumulating node weights based on SBERT similarity analysis. This approach effectively prevents the occurrence of errors due to minority discrepancies. 
\end{mybox}

\subsection{RQ4 - Usability Analysis}
Table \ref{tab:gas} presents the Gas costs of the proposed solution implemented on Ethereum. We randomly selected 10 questions and queried ChatGPT, recording the Gas consumption for each request. Specifically, \texttt{requestLLM} is the interface through which the user contract calls the oracle contract to request data, while \texttt{fulfillData} is the interface through which the oracle node returns the data. It can be observed that the Gas cost is relatively higher during contract deployment, while the subsequent calls' Gas consumption depends on the length of the queried questions and the length of the responses. This implies that integrating the capabilities of large models into the blockchain requires only a few hundred thousand Gas fees, in addition to the oracle node incentives and the LLM API service costs.

\begin{table}[h] 
\centering 
\caption{Gas cost in Ethereum.} 
\label{tab:gas} 
\resizebox{\linewidth}{!}{% 
\begin{tabular}{lccc} \hline Method & Contract Deployment & requestLLM & fulfillData \\ \hline Gas cost & 796,915 & 274,366.7 ($\pm$ 233,376.1) & 127,148.8 ($\pm$ 19,125.0) \\ \hline 
\end{tabular} } \end{table}

\begin{mybox} 
\textbf{Answer for RQ4: } The proposed method is not difficult to implement. The primary costs of the system are determined by the Gas consumption of core interface calls and the service fees associated with the LLM API. 
\end{mybox}

\section{CONCLUSION}
\label{conclusion}
This paper proposes and implements a universal framework for integrating LLMs with blockchain data, {\sysname}, aimed at overcoming interoperability barriers between blockchain and LLMs and introducing the intelligent analysis and decision-making capabilities of LLMs into smart contracts. To address the issue of heterogeneity in the data returned by nodes, we combine semantic relevance assessment with truth discovery techniques to propose a novel data aggregation method, SenteTruth. Additionally, the research constructs a dataset containing three types of questions, covering Q\&A records between 10 oracle nodes and 5 LLM models, to validate the framework's effectiveness. Experimental results demonstrate that the proposed method offers significant advantages in improving data reliability and usability.

In the future, as advancements in artificial intelligence, particularly LLMs, continue to progress, the intelligence of smart contracts is expected to undergo a revolutionary transformation. Smart contracts will no longer be confined to rigid, hard-coded rules; instead, they will be capable of autonomously making flexible decisions based on real-time conditions, the evolving needs of contract participants, and inputs from external data sources. The semantic understanding capabilities of LLMs will empower smart contracts to analyze ambiguities in contract terms, automatically identify and resolve potential conflicts or misunderstandings, and optimize trading strategies and risk management. With the deepening integration of AI and blockchain technologies, smart contracts will be able to engage in self-learning and self-adjustment, progressively enhancing the accuracy and efficiency of decision-making. This evolution will significantly broaden the applicability of smart contracts, opening new opportunities in areas such as decentralized finance, supply chain management, and smart insurance, thereby ushering in a new era of intelligent blockchain systems.

\appendix
\section*{Supplement the experimental results}
\label{append}

% 在MIX数据集下

\begin{table}[h] \centering \caption{Data accuracy under 40\% Random Response (MIX dataset)} \label{tab:1} 
\resizebox{\linewidth}{!}{ %
\begin{tabular}{llllll} \hline 
Model & ChatGPT & ChatGLM & Llama & Gemini & Hunyuan \\ \hline 
Majority Voting & 0.75 & 0.67 & 0.8 & 0.9 & 0.96 \\ 
TF-IDF Similarity & 0.6 & 0.66 & 0.99 & 0.98 & 0.9 \\ 
SBERT Similarity & 1.0 & 1.0 & 1.0 & 1.0 & 0.99 \\ 
\rowcolor{lightgray} Ours & 1.0 & 1.0 & 1.0 & 1.0 & 1.0 \\ \hline 
\end{tabular} 
} 
\end{table}

\begin{table}[h] \centering \caption{Data accuracy under 40\% Incorrect Response (MIX dataset)} \label{tab:2} \resizebox{\linewidth}{!}{% 
\begin{tabular}{llllll} \hline Method & ChatGPT & ChatGLM & Llama & Gemini & Hunyuan \\ \hline Majority Voting & 0.75 & 0.67 & 0.8 & 0.9 & 0.96 \\ 
TF-IDF Similarity & 0.6 & 0.66 & 0.99 & 0.98 & 0.9 \\
SBERT Similarity & 0.71 & 0.7 & 1.0 & 0.99 & 0.83 \\
\rowcolor{lightgray} Ours & 1.0 & 1.0 & 1.0 & 1.0 & 1.0 \\ \hline 
\end{tabular} } 
\end{table}

\begin{table}[h] \centering \caption{Data accuracy under 40\% Model Substitution (MIX dataset)} \label{tab:3} \resizebox{\linewidth}{!}{% 
\begin{tabular}{lllll} \hline 
Method & ChatGLM & Llama & Gemini & Hunyuan \\ \hline 
Majority Voting & 0.88 & 0.88 & 0.68 & 0.6 \\ 
TF-IDF Similarity & 0.81 & 0.21 & 0.25 & 0.4 \\
TF-IDF Similarity + TD & 0.99 & 0.0 & 0.0 & 0.44 \\
SBERT Similarity & 0.78 & 1.0 & 1.0 & 0.94 \\
\rowcolor{lightgray} Ours & 0.99 & 1.0 & 1.0 & 1.0 \\ \hline 
\end{tabular} } 
\end{table}

% 在PRO数据集下

\begin{table}[h] \centering \caption{Data accuracy under 40\% Random Response (PRO dataset)} \label{tab:4} 
\resizebox{\linewidth}{!}{% 
\begin{tabular}{llllll} \hline 
Model & ChatGPT & ChatGLM & Llama & Gemini & Hunyuan \\ \hline 
Majority Voting & 1.0 & 0.966 & 0.966 & 0.983 & 1.0 \\
TF-IDF Similarity & 0.833 & 0.233 & 0.466 & 0.0 & 0.133 \\ 
SBERT Similarity & 1.0 & 0.966 & 0.983 & 1.0 & 1.0 \\
\rowcolor{lightgray} Ours & 1.0 & 1.0 & 1.0 & 1.0 & 1.0 \\ \hline 
\end{tabular} } 
\end{table}

\begin{table}[h] \centering \caption{Data accuracy under 40\% Incorrect Response (PRO dataset)} 
\label{tab:5} 
\resizebox{\linewidth}{!}{% 
\begin{tabular}{llllll} \hline 
Method & ChatGPT & ChatGLM & Llama & Gemini & Hunyuan \\ \hline 
Majority Voting & 1.0 & 1.0 & 0.95 & 0.983 & 1.0 \\ 
TF-IDF Similarity & 0.833 & 0.916 & 0.783 & 0.966 & 1.0 \\ 
SBERT Similarity & 0.866 & 0.866 & 0.733 & 1.0 & 1.0 \\ 
\rowcolor{lightgray} Ours & 1.0 & 1.0 & 1.0 & 1.0 & 1.0 \\ \hline 
\end{tabular} } 
\end{table}

\begin{table}[h] \centering \caption{Data accuracy under 40\% Model Substitution (PRO dataset)} \label{tab:6} 
\resizebox{\linewidth}{!}{% 
\begin{tabular}{lllll} \hline 
Method & ChatGLM & Llama & Gemini & Hunyuan \\ \hline 
Majority Voting & 0.95 & 1.0 & 0.9 & 0.916 \\ 
TF-IDF Similarity & 0.95 & 0.966 & 0.783 & 0.933 \\ 
TF-IDF Similarity + TD & 1.0 & 1.0 & 0.983 & 1.0 \\ 
SBERT Similarity & 0.933 & 0.95 & 0.8 & 0.966 \\ 
\rowcolor{lightgray} Ours & 1.0 & 1.0 & 1.0 & 1.0 \\ \hline 
\end{tabular} 
} 
\end{table}

\section*{Acknowledgments}
The research was supported in part by the National Natural Science Foundation of China (Nos.62166004, U21A20474, 62262003), the Guangxi Science and Technology Major Project (No.AA22068070), the Basic Ability Enhancement Program for Young and Middle-aged Teachers of Guangxi (No.2022KY0057, 2023KY0062), Innovation Project of Guangxi Graduate Education (Nos. XYCBZ2024025), the Key Lab of Education Blockchain and Intelligent Technology, the Center for Applied Mathematics of Guangxi, the Guangxi "Bagui Scholar" Teams for Innovation and Research Project, the Guangxi Talent Highland Project of Big Data Intelligence and Application, the Guangxi Collaborative Center of Multisource Information Integration and Intelligent Processing.

\bibliographystyle{IEEEtran}
\bibliography{IEEEabrv,myref}

\end{document}